# Ultrafast Laser-Induced Magnetic Relaxation in Artificial Spin Ice Driven by Dipolar Interactions

D. Pecchio,[1,2] S. Sahoo,[1,2,*] O. Chubykalo-Fesenko,[3] S. Koraltan,[4] G. M. Macauley,[1,2,‡] T. Thomson,[5] D. Suess,[4] V. Scagnoli[1,2,†] and L. J. Heyderman[1,2]

[1]*Laboratory for Mesoscopic Systems, Department of Materials, ETH Zurich, 8093 Zurich, Switzerland*
[2]*PSI Center for Neutron and Muon Sciences, 5232 Villigen PSI, Switzerland*
[3]*Instituto de Ciencia de Materiales de Madrid, CSIC, Cantoblanco, 28049 Madrid, Spain*
[4]*Physik funktioneller Materialien, Universität Wien, Kolingasse 14-16, Wien, 1090, Austria*
[5]*NEST Research Group, Department of Computer Science, The University of Manchester, Oxford Road, Manchester, M13 9PL, United Kingdom*
[‡]*Present address: Department of Physics, Princeton University, Princeton, NJ 08540 USA*

**ABSTRACT**. It is of great interest to develop methods to rapidly and effectively control the magnetic configurations in artificial spin ices, which are arrangements of dipolar coupled nanomagnets that have a variety of fascinating collective magnetic phenomena associated with them. This is not only valuable in terms of acquiring fundamental understanding but is also important for future high-performance applications. Here, we demonstrate ultrafast control of magnetic relaxation in square artificial spin ice through femtosecond laser pulsed excitation, enabling rapid access to low-energy states via dipolar interactions. Time-resolved magneto-optical Kerr effect measurements reveal that, after laser-induced demagnetization, the magnetization recovers within picoseconds. During this brief transient window, dipolar coupling drives a collective magnetic ordering. Ex-situ magnetic force microscopy confirms the emergence of extended Type I vertex domains, characteristic of ground-state ordering, thus establishing ultrafast laser-driven relaxation as a route to attain the low-energy states. Through complementary energy barrier calculations and micromagnetic simulations incorporating Landau-Lifshitz-Bloch dynamics, we elucidate the underlying mechanism: transient ultrafast demagnetization followed by rapid remagnetization that enables a dipolar-driven collective rearrangement. Moreover, a tailored decreasing-fluence laser annealing protocol is shown to enhance ground-state ordering, consistently achieving over 92% ground-state vertex populations. This work opens the way to ultrafast and spatially selective control of magnetic states in artificial spin ice for spin-based computation and memory technologies, and highlights the critical interplay of thermal fluctuations, magnetostatic coupling, and transient magnetization dynamics.

## I. INTRODUCTION

The ability to manipulate the magnetic configuration of networks of nanomagnets is a key area of interest for advancing high-speed in-memory computation [1], nanomagnetic logic [2,3], neuromorphic [4,5] and reservoir computing [6], and reprogrammable magnonic crystals [7–10]. Particularly appealing for such applications are artificial spin ices (ASIs), which are arrays of magnetically coupled single-domain nanomagnets arranged on the sites of different lattices. These provide a platform to explore various collective phenomena, including magnetic frustration [11,12], relaxation processes [13], and phase transitions [14–16].

While the field has significantly expanded, now encompassing a range of different lattices and nanomagnet shapes [17], ASIs were originally conceived to mimic the behavior of their crystal spin ice counterparts: the rare-earth pyrochlores [18]. Here, stadium-shaped nanomagnets were arranged on the square or kagome lattices [11,19], with the pronounced shape anisotropy associated with their elongated shape confining their overall magnetic moments – or macrospins – to point in one of two stable directions parallel to the nanomagnet long axis, resulting in Ising-like behavior. Importantly, in contrast to their crystal counterparts, ASIs allow direct real-space imaging of the magnetic configuration, enabling the observation of individual reversal events and collective dynamics with magnetic imaging techniques [19,20].

[*]Contact author: sourav.sahoo@psi.ch
[†]Contact author: valerio.scagnoli@psi.ch

Control of the magnetic configuration, and hence energy, of ASIs is a key goal, with methods to allow magnetic relaxation being progressively refined [13,21–28]. With each advancement, the aim has been to achieve low-energy states more effectively, often as a means to access the ground state. The methods employed for this include demagnetization with a magnetic field [21,22], thermal annealing [13,23–26] and even patterning nanomagnets thin enough that they are thermally active close to room temperature [20,27]. However, while these techniques are effective in terms of bringing the system to a lower energy configuration, they suffer from limitations in spatial control and operational speed [13,28]. To overcome these limitations, various light-induced methods for controlling the magnetization of the nanomagnet ensemble have been developed, including continuous-wave all-optical magnetic switching [29,30] and plasmon-assisted photo-heating [1,31]. Although these techniques allow specific regions within an ASI to be addressed, they do not represent the fastest means of manipulating the magnetization, as they typically operate on timescales ranging from several nanoseconds to milliseconds. Achieving magnetization control on the sub-nanosecond timescale is particularly compelling, as it matches the characteristic timescale (GHz frequencies) of modern computing hardware, positioning ASIs as viable candidates for integration into high-speed, spin-based logic and memory technologies. In this context, laser-driven ultrafast demagnetization [32] with femtosecond pulses has emerged as one of the most promising methods for achieving magnetization control on sub-nanosecond timescales. However, so far, femtosecond laser experiments have been primarily employed to study ultrafast dynamics and all-optical switching in ferromagnetic and ferrimagnetic thin films, as well as in uncoupled magnetic nanostructures [33]. This raises the question of whether femtosecond-laser-induced magnetization dynamics could be harnessed to rapidly drive ASIs to their lowest energy state.

Here, we demonstrate that femtosecond laser excitation, as illustrated in Fig. 1(a), drives rapid magnetic relaxation of artificial square ices, enabling faster access to the ground state compared with existing methods [13,21–31]. Our results show that this relaxation process, governed by the magnetostatic (or dipolar) interactions between the nanomagnets, occurs on the same picosecond timescale as the magnetization recovery within each nanomagnet following a laser pulse. This is unexpected, as dipolar-driven collective relaxation is typically considered to be much slower [34]. Nevertheless, under these conditions, the system can reach a low-energy state in the exposed region. Our approach thus provides an order of magnitude increase in speed, while simultaneously offering spatial control over the resulting magnetic configuration, with local excitation of micrometer-scale regions not possible with global thermal or magnetic-field-driven demagnetization protocols.

To better understand the mechanism of such fast relaxation of the artificial square ice, we employed a combination of simplified and improved string method (SISM) energy-barrier calculations and Landau-Lifshitz-Bloch (LLB) micromagnetic simulations to model ultrafast demagnetization-remagnetization dynamics. These different simulation approaches each span different time and temperature scales, highlighting the critical relationship between the initial ultrafast demagnetization and the subsequent partial recovery of the magnetization within individual nanomagnets. We propose that the magnetization within nanomagnets must be sufficiently reduced on a short timescale to allow macrospin reorientation, while also recovering sufficiently quickly to re-establish dipolar interactions between nanomagnets in the artificial square ice. This transient regime, in which thermal fluctuations and magnetostatic interactions coexist, facilitates collective relaxation into low-energy configurations, as directly evidenced by the experimental observation of long-range magnetic order.

## II. SAMPLE PREPARATION AND EXPERIMENTAL SETUP

Our samples consist of arrays of thermally evaporated 8 nm-thick permalloy ($Ni_{80}Fe_{20}$) nanomagnets on a silicon (100) substrate covered with a native oxide layer (∼1-2 nm thick) [Fig. 1(b)], patterned with electron beam lithography. These artificial square ices have dimensions of 200 μm × 200 μm and the nanomagnets are of length $L$ = 300 nm and width $W$ = 100 nm, and the square lattice constant is $a$ = 400 nm. The nanomagnets are capped with 2 nm of aluminum to prevent oxidation, and the smallest gap between the nanomagnets is $g$ ≈ 15 nm.

First, longitudinal time-resolved magneto-optical Kerr effect (TR-MOKE) measurements were performed using a Nd:YAG pulsed fiber laser in pump-probe configuration, with a 200 kHz repetition rate, 1030 nm fundamental wavelength and a nominal pulse duration of 80 fs. These measurements were used to record the time evolution of the magnetization dynamics of the sample following in-field ultrafast laser excitation [Fig. 1(d)]. S-polarized probe pulses were derived from the fundamental wavelength, while the P-polarized pump (excitation) pulses with a wavelength of 515 nm were obtained via second-harmonic generation using a barium borate (BBO) crystal. During TR-MOKE measurements, the average power of the pump beam was 5 mW (0.88 mJ/cm²), while the power of the probe beam was 300 μW (0.05 mJ/cm²). Orthogonal polarizations of the pump and probe beams were used to minimize optical interference and enable selective detection of the Kerr rotation signal. To effectively distinguish the magnetic signal from the background noise, we performed these measurements with a magnetic field of $H$ = 500 Oe applied along both the $[11]$ and $[\bar{1}\bar{1}]$ directions for every data point. The Kerr signal shown in Fig. 1(d) represents the normalized change in Kerr rotation measured during this process, $\Delta\theta_K^{norm}(t) = \frac{\theta_K(t) - \theta_K^{eq}}{\theta_K^{eq} - \theta_K^{min}}$, where $\theta_K(t)$ is the Kerr rotation measured at a pump-probe delay $t$, $\theta_K^{eq}$ is the equilibrium Kerr signal before excitation (i.e., for $t < 0$), and $\theta_K^{min}$ is the minimum value of the Kerr signal reached after excitation, corresponding to the maximum degree of demagnetization. This normalization ensures that the signal is zero before excitation and reaches -1 at the point of maximum demagnetization. Although the longitudinal Kerr rotation is not a direct measure of magnetization in absolute units, it is, to first approximation, linearly proportional to the in-plane component of the magnetization in the probed volume. At equilibrium, the Kerr signal is set to zero to remove the static offset and highlight dynamic changes induced by the pump pulse. As shown in Fig. 1(d), the TR-MOKE signal exhibits a three-stage response: an ultrafast demagnetization on the sub-picosecond timescale, followed by a fast partial recovery and then a slower complete recovery with characteristic time constants of ∼30 ps and ∼300 ps, respectively.

This initial TR-MOKE characterization of the in-field magnetization dynamics, where sample heating is minimal, serves as a reference point that allows the effects of laser exposure on the magnetic configuration of the ASIs to be interpreted as it evolves following the excitation. The ASIs were exposed with pump pulses with a pulse duration of 100 fs at the sample position, an incidence angle of 45°, a full width at half maximum (FWHM) laser spot size of 85 μm, and a laser fluence between 6.1 mJ/cm$^2$ and 9.7 mJ/cm$^2$ (corresponding to an average laser power between 35 mW and 55 mW), in the absence of an external magnetic field, as illustrated in Fig. 1(a). All experiments were performed under ambient conditions.

The resulting magnetic configurations of the artificial square ices were subsequently imaged via ex-situ magnetic force microscopy (MFM).

### III. ULTRAFAST LASER-INDUCED MAGNETIC RELAXATION IN SQUARE ASI

A convenient framework to describe the magnetic state of an artificial square ice is to consider the vertex populations. In the artificial square ice, four neighboring nanomagnets surround a lattice vertex in a cross-like arrangement, which is highlighted in red in Fig. 1(b). Four distinct vertex types, Type I to Type IV [Fig. 1(c)], with increasing dipolar energy, can be identified based on the orientation of the macrospins of the constituent nanomagnets. Although the total energy of the system arises from the full set of dipolar interactions, the relative populations of the different vertex types serve as an effective proxy for the energy of the system: the more low-energy vertex configurations that are present in the system, the closer the system is to its ground state. The lower energy Type I and Type II vertex configurations are both characterized by two macrospins pointing towards the center of a vertex and two pointing away, following the so-called "ice rule", similar to that observed in a bulk crystal spin ice [18]. However, Type I vertex configurations have a lower energy than Type II vertex configurations because the dipolar interactions between the pairs of nanomagnets at a vertex depend on their relative orientations and separation, which are different for nearest neighbor orthogonal nanomagnets and the parallel nanomagnets across the vertex [11,35]. As a consequence, the ground state of an artificial square ice consists of an alternating tiling of the two Type I vertex configurations, and this ground state is doubly degenerate. An array in which all vertices adopt the same Type II vertex configuration can be easily obtained by applying a magnetic field along any of the four ⟨11⟩ directions of the artificial square ice. Type III vertex configurations have a higher dipolar energy than Type II vertex configurations, with three macrospins pointing in and one pointing out, or vice versa. Type III vertex configurations typically appear as defects within the chains of Type II vertices that form at the boundaries between ground state domains [36]. Finally, Type IV vertex configurations are the highest in energy, with all the macrospins pointing in or all pointing out but are usually not observed.

To monitor the extent to which laser excitation drives artificial square ice into low-energy configurations, several of 200 μm × 200 μm nanomagnet arrays are prepared in a well-defined configuration. This is achieved by applying a magnetic field significantly higher than the coercive field of the nanomagnets along the [$\bar{1}\bar{1}$] direction (i.e., opposite to [11]). The field is then removed, so that the macrospins of the horizontal (vertical) nanomagnets shown in Fig. 1(b) point left (down). As a result, all vertices are in the same Type II configuration indicated with a black circle in Fig. 1(c). Subsequently, different copies of the same artificial square ice were exposed to a train of 100 fs laser pulses for 1 s at a repetition rate of 200 kHz (i.e., one pulse every 5 μs, with the laser effectively extinguished between pulses) with a given laser fluence. Each artificial square ice was exposed to a different, but constant, laser fluence, which was increased from 6.1 mJ/cm$^2$ to 9.7 mJ/cm$^2$ in steps of 0.9 mJ/cm$^2$. Subsequently, the magnetic configuration of each artificial square ice was imaged using MFM over an area of 20 μm × 20 μm, which contains approximately 5000 nanomagnets. An example of a resulting MFM image is provided in Fig. 2(a), where each nanomagnet is associated with a bright and a dark contrast that results from the stray field associated with opposite magnetic poles at the ends of the nanomagnets. This contrast allows us to determine the orientation of the macrospin assigned to each nanomagnet. From this, the vertex configurations of the region are mapped as shown in Fig. 2(b), with each vertex type assigned a different color and symbol. Here, the two degenerate ground state domains consisting of ordered Type I vertices are indicated by green and purple dots. Vertices in a Type II configuration are identified by a black arrow that indicates the direction of the net magnetic moment. Similarly, the net magnetic moment of vertices in a Type III configuration is indicated with white arrows, enclosed in red and blue dots to denote positive (three-in, one-out) and negative (one-in, three-out) magnetic charges, respectively. Here we are referring to the effective magnetic charge defined in the dumbbell model, where each nanomagnet is treated as a dumbbell with opposite charges at its ends [19,37]. The magnetic charge at a vertex is then the net charge associated with the four dumbbell charges residing at the vertex.

As mentioned above, we began by applying trains of laser pulses of different but constant fluence. At a fluence of 6.1 mJ/cm$^2$, we observed no change in the artificial square ice. However, when we employed a fluence of 7.0 mJ/cm$^2$, we saw reversal of macrospins in the central region of the scanned area, while the vertices near the edges remained in their initial Type II configuration. Since the laser pulses have a Gaussian energy distribution over the sample surface, reversal was only observed in the central region where the fluence was higher than the threshold required to reverse the macrospin within nanomagnets. In the central region, we observe the laser-induced formation of Type I domains, reflected by the green and purple regions in the vertex configuration map in Fig. 2(b). The size of these domains remains relatively small, spanning only a few vertices in width. We ascribe this observation to the fact that the laser fluence is only slightly above the switching energy threshold, providing just enough energy to induce local switching but not sufficient energy to drive the formation of extended Type I domains. The vertex configuration map in Fig. 2(b) also illustrates how Type II vertex chains serve as boundaries separating Type I domains. Finally, we also observe vertices in a Type III configuration. These are primarily located at the interfaces between the unswitched Type II region and the newly formed Type I domains. Although these exposures were performed using 1 second - long pulse trains at 200 kHz to ensure stable excitation conditions, control experiments confirmed that magnetization reversal and Type I domain formation also occurs with a single femtosecond pulse, provided that the fluence is sufficiently high (details of the single-pulse experiments are given in Section I of the Supplemental Material [SM]).

As the laser fluence was increased to 7.9 mJ/cm$^2$ [Fig. 2(c), where Type I symbols are replaced with squares of the same color to highlight domains with opposite nanomagnet polarization, while Type II and Type III symbols are omitted], we observed that the magnetic moments in the entire 20 μm × 20 μm area scanned with MFM received sufficient energy to reverse. In addition, the size of the Type I domains increased, indicating that more nanomagnets could undergo reversal, and a lower-energy magnetic configuration was reached. Increasing the fluence to 8.8 mJ/cm² [Fig. 2(d)] promotes the formation of even larger Type I domains, bringing the system closer to its ground state, while a higher fluence of 9.7 mJ/cm² [Fig. 2(e)] did not appear to change the domain sizes significantly. In order to quantify this, we determined the populations of different vertex types, which allows us to gauge the degree of order in the system, and track their evolution with laser fluence, as shown in Fig. 2(f) for the three observed vertex types. As the fluence is increased, the Type I population increases, until it reaches approximately 80% at 8.8 mJ/cm². Increasing the fluence still further produces no substantial increase in the Type I population. The Type II vertices, found in the domain boundaries, are significantly reduced to 20%, while the limited number of Type III vertices observed at low fluences remain close to zero. These vertex populations are comparable to those achieved through thermal annealing protocols, where similar Type I vertex populations are observed under optimal conditions [23,26]. However, we have achieved this on a significantly faster timescale.

In summary, when sufficient energy is delivered to the artificial square ice via a femtosecond laser pulse, a significant macrospin dynamics is triggered that allows the system to reorganize into a lower-energy state. Given the ultrashort laser-sample interaction time (<200 fs) and the fact that magnetization recovery within nanomagnets completes within 2 ns, as seen in Fig. 1(d), our results raise the question of when and how macrospin ordering occurs across the artificial square ice as a whole. Notably, our TR-MOKE measurements show that the system recovers approximately 60% of its original magnetization within just 40 ps following laser-induced demagnetization. This suggests that the macrospin rearrangement should entail sub-nanosecond magnetic fluctuations, governed by dipolar coupling interactions acting on comparable timescales.

Ideally, time-resolved imaging experiments [38] would be used to track the evolution of the artificial square ice configuration following laser excitation. However, real-space time-resolved imaging with techniques such as free-electron laser-based time-resolved XMCD-PEEM is highly challenging. In particular, these methods suffer from limited magnetic contrast and low signal strength under typical stroboscopic conditions, which would be exacerbated by the fact that the nanomagnets have in-plane magnetization and are only 8 nm-thick. Given these constraints, we turned to an alternative approach: by systematically modifying the annealing protocol, we sought indirect insight into the underlying relaxation dynamics.

For this, a protocol based on exciting the system with a series of laser pulses with a gradually decreasing fluence over time (exposure duration of 1 s), from 9.7 mJ/cm$^2$ down to 6.1 mJ/cm$^2$ in steps of 13.2 nJ/cm$^2$ (corresponding to 273 steps with ~3.7 ms per step), was implemented and this was found to significantly improve the degree of magnetic ordering compared to exposures of a constant fluence. While we did not systematically explore other protocols or step sizes, this fine decrement was chosen to ensure a gradual reduction in the energy, allowing controlled, gradual ordering of the magnetic moments. Conceptually, this strategy resembles simulated annealing [39], in which a gradual reduction in energy input enables the system to escape local energy minima and progressively evolve towards more ordered configurations. The protocol was found to be successful based on its consistent ability to produce large Type I domains with minimal defects across multiple samples. To assess the effectiveness and reproducibility of this approach, we applied the protocol to four different but nominally identical artificial square ices. The resulting vertex configuration map for one of these is shown in Fig. 3(a), while Fig. 3(b) displays the populations of Type I, II and III vertices in each of the four samples following this decreasing fluence protocol. Notably, we achieve an average Type I vertex population of 92.4 ± 0.6 %.

To understand how large Type I domains form in artificial square ice with this decreasing-fluence laser excitation protocol, we first need to acknowledge that there is a distribution of the energy barriers for nanomagnet switching [19]. This can arise from defects and other variations in the microstructure of the permalloy introduced during fabrication. At the highest laser fluences, all macrospins can reorient. However, as the laser fluence, and thus the energy delivered per pulse, decreases, the probability that the magnetic moment of a given nanomagnet will switch becomes strongly dependent on its energy barrier. The nanomagnets with the highest energy barriers will become effectively blocked on the timescale of the laser excitation, meaning that their switching probability is drastically reduced. These nanomagnets can then act as nucleation sites for the formation of Type I vertex domains. Here, the stray field generated by such nanomagnets with "frozen" macrospins will bias the orientation of the macrospins in neighboring nanomagnets to form energetically favorable vertex configurations, so that small Type I vertex clusters form that will grow in size. Crucially, once a Type I domain forms, its magnetostatic environment stabilizes it. This means that, as the Type I domain grows, the probability that nanomagnets will switch back into a higher-energy state becomes infinitesimally small, as the energy barrier for such transitions increases. In contrast, nanomagnets at the domain boundaries can still undergo reversal, promoting further expansion of the domain into adjacent higher-energy regions. This self-reinforcing process continues until expanding Type I domains collide, forming boundaries composed of Type II vertex configurations between them. If there is sufficient time, these Type II boundaries can then shrink and annihilate through the motion of Type III vertices [36]. The formation of large, well-ordered Type I domains shown in Fig. 3(a), therefore suggests a collective relaxation mechanism governed not only by stochastic nucleation but also by the local energetics that drive the growth and coalescence of Type I domains.

## IV. MICROMAGNETIC SIMULATIONS OF MAGNETIZATION DYNAMICS IN SQUARE ASI

We now clarify how long-range magnetic order can emerge following ultrafast laser excitation within the rapid transient window in which the magnetization within individual nanomagnets is still recovering. Our observations suggest that dipolar interactions play a key role during the picosecond-timescale magnetization recovery that comes after its ultrafast (sub-picosecond) quenching of the magnetization. Yet, it remains unclear how these interactions become effective so rapidly. To clarify this, we performed numerical simulations designed to reproduce the physical conditions under which nanomagnet relaxation occurs and to shed light on the mechanisms governing their fast collective response. In a first step, we calculated the energy barrier for a transition from a Type II to a Type I vertex configuration as a function of temperature [Fig. 4(a)], using the simplified and improved string method (SISM) [40–43]. To account for temperature effects, the saturation magnetization and exchange stiffness were rescaled according to the temperature [40].

A detailed view of the entire transition pathway, comprising the evolution of the energy profile along the discretized trajectory at different temperatures, is provided in Fig. S2 [SM], with details of the modeling of the magnetization reversal and associated magnetic parameters are given in Section II of the Supplemental Material [SM]. For this, we modeled the simultaneous reversal of two nanomagnets to give a direct transition from the initial Type II to the final Type I vertex configuration [Fig. S3], which is characterized by a single energy barrier with no intermediate local minima. This corresponds to a minimum energy path in which both nanomagnets switch together without forming metastable states such as Type III vertex configurations. This contrasts with the mechanism that is expected to occur during thermal annealing where the transition from a Type II to a Type I vertex configuration would proceed through sequential switching of two coupled nanomagnets per vertex, so that it passes through an intermediate Type III vertex and involves two distinct energy barriers. Our choice of reversal mechanism is guided by the experimental observation that magnetization reversal can be triggered by a single ultrashort laser pulse. While the exact sequence of switching events remains unclear, assuming simultaneous reversal offers a reasonable approximation within the ultrafast timescale under consideration.

Our calculations reveal that, although the energy barrier separating vertices in Type II and Type I configurations decreases with increasing temperature [Fig. 4(a)], it remains higher than 20 eV per nanomagnet at 750 K, which is still several orders of magnitude above $k_BT$ (~66 meV at this temperature). Achieving a significant probability of thermally activated switching would therefore require the system to remain at such a high temperature for much longer than is typically achieved through ultrafast excitation with femtosecond lasers. This rules out thermally activated switching (i.e., the superparamagnetic regime) as the dominant mechanism during single-pulse excitation. Instead, our results support the interpretation that magnetization reversal occurs due to ultrafast laser-induced demagnetization, followed by rapid magnetization recovery and dipolar-driven reorientation, a process that takes place only when the system is transiently driven extremely close to the Curie temperature (770 K). It is plausible that this condition is met during the short time window following laser excitation – between ultrafast demagnetization and fast remagnetization [Fig. 1(d)] – when the electron temperature remains very close to the Curie temperature on a sub-picosecond timescale. We believe this transient state allows for an initial disordering of the magnetization and creates favorable conditions for dipolar interactions to drive the system toward a lower-energy configuration.

To complement the SISM analysis, we employed Landau-Lifshitz-Bloch (LLB) micromagnetic simulations [44–46] coupled to the two-temperature model [47,48], which are specifically designed to capture ultrafast demagnetization dynamics and incorporate longitudinal relaxation, i.e., the temperature-dependent changes in the modulus of the magnetization vector $|M|$, in contrast to standard micromagnetic models where the $|M|$ remains fixed. Because it allows $|M|$ to vary with temperature and time, the LLB simulations are appropriate for modelling the magnetization dynamics following ultrafast laser excitation of artificial square ice.

The simulations were performed with a time step of 1 femtosecond, using magnetic parameters consistent with those adopted in the SISM calculations (Supplemental Material, Section II [SM]), and assuming a pulse duration of 60 fs. The heating effect of the laser excitation is introduced via the two-temperature model, using parameters listed in Section III of the Supplemental Material [SM]. These parameters define the temporal evolution of the electron and phonon temperatures following the laser excitation and are selected to qualitatively reproduce the conditions under which magnetization reversal is expected to occur. However, since the actual laser absorption in the experimental samples is not precisely known, the simulated fluence is not intended to correspond quantitatively to the experimental values, and should be regarded as a fitting parameter that ensures demagnetization in the model. As can be seen in Fig. 4(b), under these conditions, the two-temperature model shows a rapid rise in electron temperature $T_e$ – on the order of a hundred femtoseconds – when the laser pulse transfers energy to the magnetic material, followed by thermalization between the phonon and electron baths. The magnetization dynamics are coupled to the electron temperature via spin-orbit interaction. However, because spin relaxation has a finite timescale (≈ 100 fs), the spin system does not immediately follow the electron temperature. As a result, the average magnetization modulus $\langle|M|\rangle$ reflects the spin temperature, which lags behind the electron temperature on a sub-picosecond timescale.

To explore ultrafast ASI dynamics further, we used the same LLB code to simulate the temporal and spatial evolution of the magnetization for two unit cells of artificial square ice with periodic boundary conditions [Fig. 4(c)]. Further details about the simulation procedure and parameters are provided in Section III of the Supplemental Material [SM]. Starting from a state in which all vertices are in a Type II configuration at $t = 0$, following ultrafast laser excitation, each nanomagnet undergoes full demagnetization, which is achievable only when the electron temperature significantly exceeds the Curie temperature. As a result, each nanomagnet is in a disordered state, with fluctuating, uncorrelated spins on the atomic level, and therefore no net magnetization. During the subsequent recovery, magnetic domains nucleate within the nanomagnets themselves, and subsequently grow and coalesce, reaching about 30 nm in size at $t$ = 100 ps. On this timescale, the intrinsic magnetization is almost fully recovered, and magnetostatic interactions become active. At a later time, $t$ = 3 ns, the magnetic configuration within the nanomagnets is either single-domain or hosts a vortex-like structure, which is a metastable configuration. As the system evolves, each vertex may relax into a low-energy Type I configuration via dipolar interactions. However, due to the high computational cost of LLB simulations, we are limited to small systems and cannot run enough statistically independent realizations to assess whether the system consistently relaxes into a lower-energy state, as experimentally observed.

To better understand the role of magnetic order during the relaxation process, we monitor both the average magnetization modulus, $\langle |M| \rangle$, and the average component of the magnetization along the x-direction, $\langle M_x \rangle$ [see coordinate axes in Fig. 4(c)]. While $\langle |M| \rangle$ tracks the evolution of the spin temperature and the degree of local demagnetization, $\langle M_x \rangle$ is sensitive to the net orientation of the macrospins along the x-axis and thus reflects the global magnetic configuration. This distinction is crucial in artificial square ice, where the Type II state exhibits a net magnetization along x (ferromagnetic-like), whereas the Type I ground state has zero net magnetization (antiferromagnetic-like). Therefore, tracking $\langle M_x \rangle$ allows us to distinguish whether the system returns to the original Type II state or evolves towards a Type I configuration. However, relying exclusively on $\langle M_x \rangle$ may lead to ambiguities, as certain Type III and Type IV vertex configurations also exhibit vanishing in-plane magnetization and might be mistakenly identified as Type I. In practice, this issue does not arise in the LLB simulations, as these high-energy configurations are never observed during the recovery process. This is likely to be because the simulations are restricted to only two unit cells of artificial square ice with periodic boundary conditions, which significantly limits the number of accessible vertex configurations.

In Figure 4(b), we show the temporal evolution at high fluence, where the system does not recover its initial $\langle M_x \rangle$ value, indicating a transition from Type II to Type I vertex configurations. In contrast, for low-fluence, as shown in Fig. S4 [SM], $\langle M_x \rangle$ recovers after the pulse, consistent with a reversible excitation of the Type II configuration and high energy barriers. These two scenarios highlight that it is only when the energy delivered to the electronic system is sufficient to raise its temperature above the Curie temperature and fully demagnetize the nanomagnets that the system can access new microstates and, potentially, undergo a reconfiguration of its magnetic state. Note that, in the LLB simulations, although the intrinsic magnetization recovers on an ultrafast timescale, some elements transiently form vortex-like states, which require a longer time – on the order of nanoseconds – to become single domain.

From the LLB simulations, we observe that a single laser pulse can induce a transition from Type II to Type I vertex configurations. However, preliminary results indicate that the final outcome may not be fully deterministic, since simulations under identical conditions can yield different magnetic states due to thermal fluctuations and the influence of local dipolar fields. While LLB simulations are limited to single-pulse events and small system sizes, theoretical considerations based on kinetic Monte Carlo [49,50] and master equation [51] approaches suggest that even large energy barriers can be overcome via the cumulative effect of many laser pulses. This supports the interpretation that multiple excitations are needed to reliably drive the system towards its ground state, with dipolar interactions playing a key role in guiding the relaxation.

Our results demonstrate that laser excitation enables the system to overcome significant energy barriers, and that long-range magnetic order can emerge within sub-nanosecond timescales. These observations suggest that the relaxation process may proceed via two distinct pathways. In the first scenario, the nanomagnets transiently demagnetize and recover their magnetization, temporarily remaining in a multidomain or vortex-like state during the recovery. As the magnetization begins to recover, each nanomagnet may develop a net magnetic moment oriented along its long axis, even if the internal magnetization texture is not yet fully uniform. This partial recovery means that the average magnetization of each nanomagnet can still be thermally reconfigurable. In this case, dipolar fields from neighboring elements can influence the orientation of these average moments, allowing the system to collectively evolve towards an ordered configuration before the nanomagnets become magnetically frozen. This mechanism suggests that full monodomain formation is not strictly required to achieve the ground state, as long as thermal fluctuations and dipolar interactions act in concert during the recovery. However, in this scenario, multidomain or vortex-like configurations will tend to inhibit dipolar-driven reorganization and increase the likelihood of the system being trapped in higher-energy configurations.

Another, perhaps more plausible, relaxation pathway is one in which the nanomagnets recover into quasi-macrospin states with a sufficient uniform magnetization but also sufficiently reduced energy barriers – due to the high temperature and reduced saturation magnetization – allowing thermally activated switching within the transient window. In this regime, the system may follow an Arrhenius-like behavior, where the effective switching time is governed by the exponential dependence on the energy barrier, provided that the attempt rate $\tau_0 \approx 10^{-12}$ s [20,52,53] and the temperature remains sufficiently close to the Curie temperature. Since the system remains near the Curie temperature for a very short time, on the order of 1.2 picoseconds, the probability of such a reversal is very low. However, if a large enough fraction of nanomagnets undergoes thermally-induced macrospin reversal before the system cools down, the dipolar interactions can guide the ensemble into a low-energy Type I configuration at nearly every vertex. This process should be more efficient with multiple laser pulses. We note here that the sequential freezing process, described earlier in the context of the experimental results (Section III) and arising from the distribution of energy barriers, is likely to play a role in both of these relaxation pathways, particularly in the multi-pulse regime where the system undergoes progressive reordering.

Crucially, both mechanisms rely on a delicate interplay between the degree of demagnetization, the rate of remagnetization, and the strength of dipolar interactions. Despite the simplifications of the LLB model and the finite size of our simulations, the robustness and reproducibility of the experimental results suggest that this ultrafast access to the ground state is not coincidental but reflects an intrinsic and effective pathway for energy minimization.

## V. CONCLUSIONS

In summary, we have shown that femtosecond laser excitation can drive magnetic relaxation in artificial square ices on sub-nanosecond timescales, enabling the system to access low-energy configurations and reach the ground state. While our main experiments employ pulse trains for stability, control measurements confirm that a single ultrashort pulse is sufficient to trigger magnetization reversal and initiate the relaxation process toward a more ordered magnetic configuration, provided the fluence is high enough. The magnetic relaxation relies on the delicate interplay between ultrafast demagnetization and partial recovery of the magnetization, allowing dipolar interactions to drive collective ordering. By applying a decreasing-fluence annealing protocol, we further enhanced the effectiveness of the relaxation, achieving a population of the lowest-energy Type I vertices exceeding 92%. We have thus demonstrated that the transient high-temperature regime provides favorable conditions for dipolar-driven reordering, offering a rapid and spatially selective route for controlling magnetic configurations in ASIs, providing a promising platform for ultrafast spin-based computing [5].

Although LLB micromagnetic simulations reproduce the transient magnetization dynamics well, they suggest that relaxation to the lowest-energy state after a single excitation may not be fully deterministic, due to the influence of thermal fluctuations and dipolar interactions. Moreover, it is not viable to model the cumulative effects of multiple laser pulse excitations, as each excitation starts from the same initial state, and the computational cost means that simulations over long timescales, or for large ensembles, are not feasible. This highlights current limitations in modeling the subtle interplay between demagnetization, thermal fluctuations, and dipolar coupling on longer timescales. Bridging this gap will require multiscale approaches including atomistic spin dynamics [45,54] and kinetic Monte Carlo techniques [13] to fully capture the mechanisms behind ultrafast relaxation and ordering in arrays of coupled nanomagnets, as exemplified by recent multiscale modeling of magnon dynamics and spin-current-driven reordering in ferrimagnets [55].

Finally, while our work has focused on artificial square ice, the ultrafast relaxation mechanisms revealed here may also play a role in other artificial spin ice geometries, including both two-dimensional systems [17] and recently developed three-dimensional systems [56], offering exciting prospects for future investigations.

## ACKNOWLEDGMENTS

We would like to thank Peter Derlet for theoretical discussions, Matteo Savoini for technical support with the laser setup, and Anja Weber for technical assistance with nanofabrication. This work was supported by the Swiss National Science Foundation (Grant No. 200020_200332).

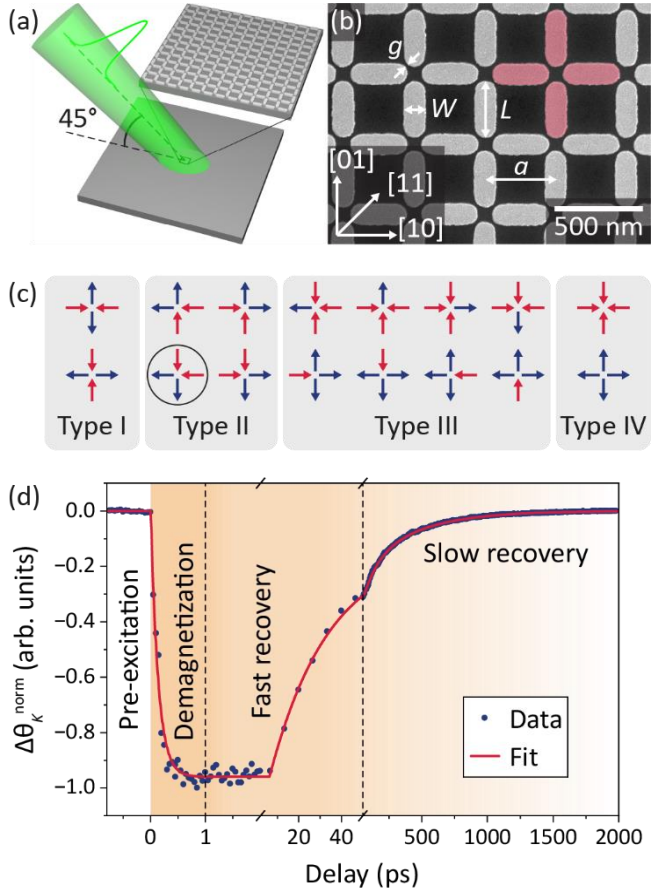

FIG. 1. Applying ultrafast pulsed laser excitation to an artificial square ice. (a) Schematic of the pulsed laser experimental geometry and sample layout. (b) Scanning electron microscope image of the artificial square ice ($L$ = 300 nm, $W$ = 100 nm, $a$ = 400 nm, $g \approx$ 15 nm), with a four-nanomagnet vertex highlighted in red. (c) Vertex types (Type I–IV) ordered in terms of increasing dipolar energy. Red (blue) arrows indicate macrospins pointing towards (away from) the vertex. The initial state of the artificial square ice used in the laser annealing experiments has all vertices in the Type II vertex configuration highlighted with a black circle. (d) Time-resolved changes of Kerr rotation representing the temporal evolution of the magnetization of the artificial square ice after pulsed laser excitation. The signal is normalized such that the Kerr rotation is zero at equilibrium and –1 at the point of maximum demagnetization. The vertical dashed lines demarcate the characteristic temporal regions: demagnetization, fast recovery, and slow recovery.

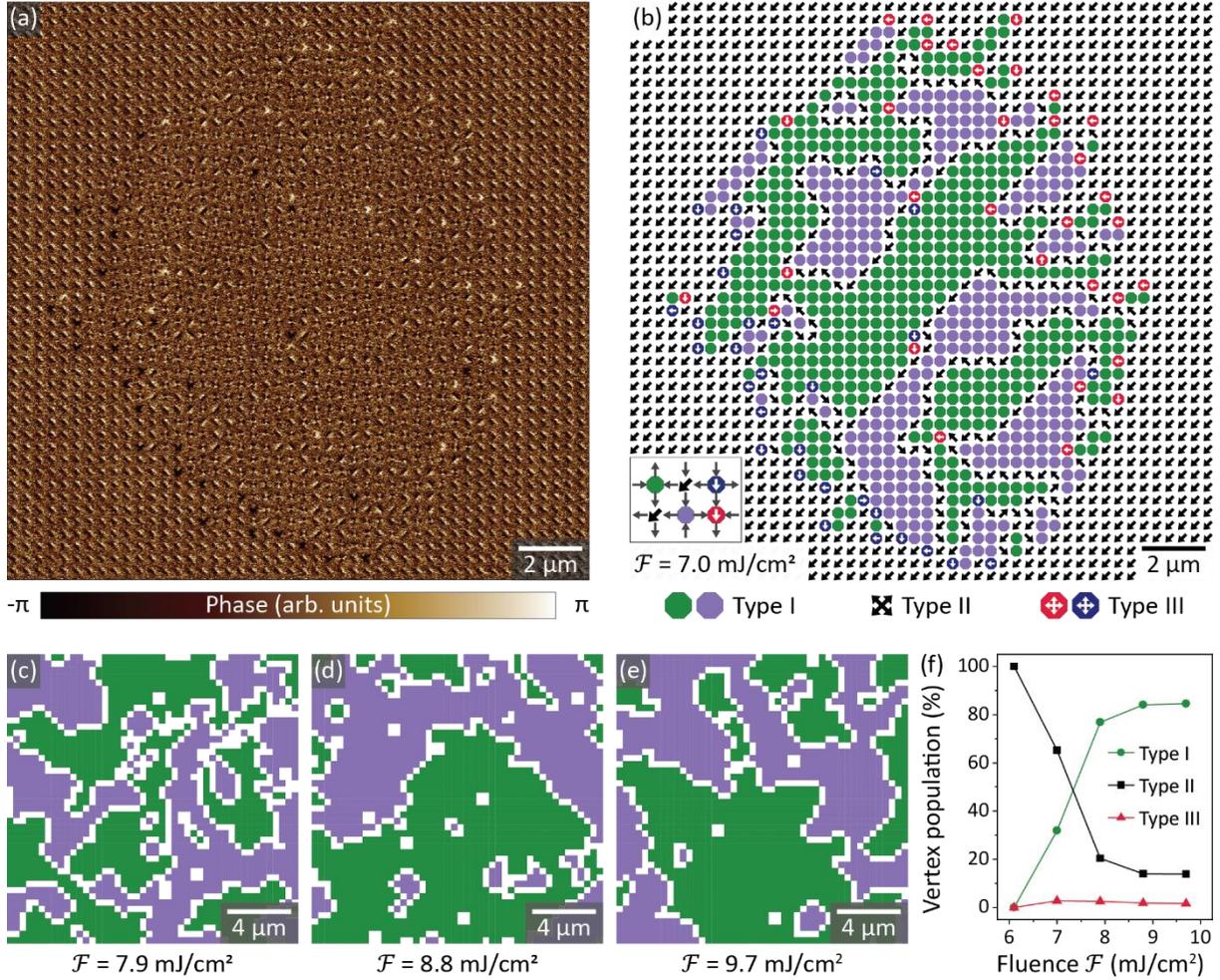

FIG. 2. Vertex configurations and populations in artificial square ice following laser excitation with a constant fluence over time. (a) Magnetic force microscope phase contrast image and (b) corresponding map of the vertex configurations following laser exposure with a laser fluence of $F = 7.0$ mJ/cm$^2$. The different vertex types are color coded with symbols as indicated. Simplified maps of the vertex configurations are provided for the magnetic configurations of the artificial square ices exposed to fluences of (c) 7.9 mJ/cm$^2$, (d) 8.8 mJ/cm$^2$ and (e) 9.7 mJ/cm$^2$. Here, Type I symbols are replaced with squares of the same color to highlight the Type I domains; Type II and Type III symbols are omitted. (f) Populations of Type I, Type II and Type III vertices as a function of laser fluence $F$ extracted from the maps of the vertex configurations.

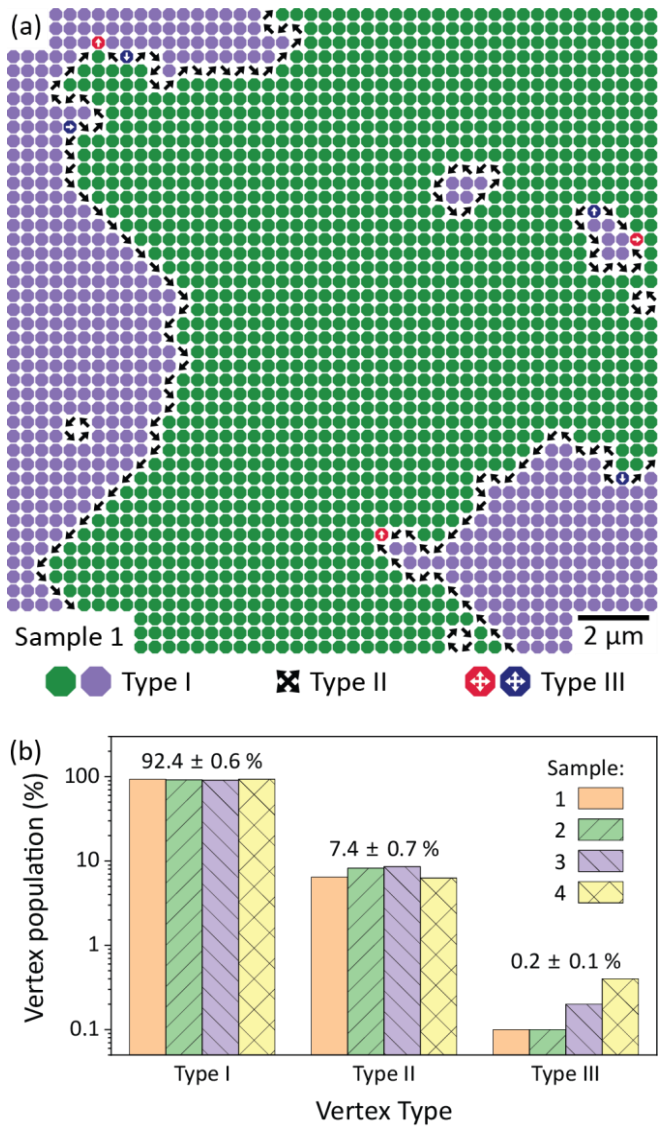

FIG. 3. Vertex configurations and populations following laser excitation with a series of laser pulses of a gradually decreasing fluence. Here four different but nominally identical artificial square ices were exposed to a laser of decreasing fluence over time (exposure duration of 1 s) going from 9.7 mJ/cm$^2$ down to 6.1 mJ/cm$^2$ in steps of 13.2 nJ/cm$^2$. (a) Map of the vertex configurations of an artificial square ice (Sample 1). (b) Type I, Type II and Type III vertex populations extracted from the MFM images for Samples 1 to 4. The average vertex population along with its standard deviation for each vertex type is given above the graphs and the corresponding vertex configuration maps are given in Fig. S1 [SM].

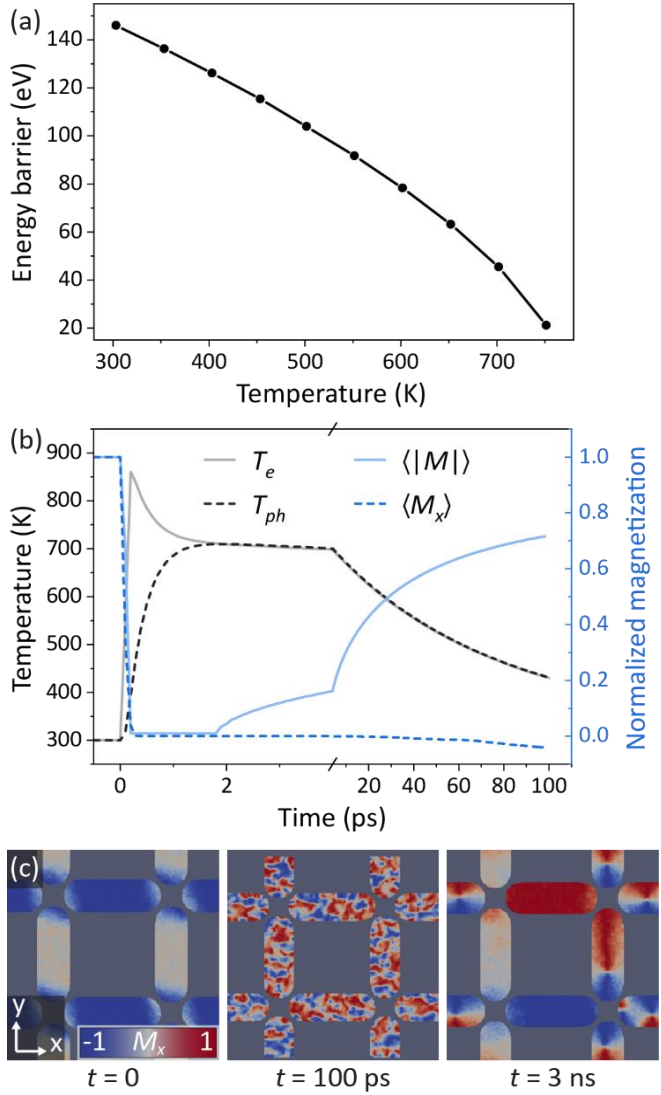

FIG. 4. Micromagnetic simulations of energy barrier and ultrafast magnetization dynamics. (a) Calculated energy barrier to switching from a Type II to a Type I vertex configuration as a function of temperature, using the Simplified and Improved String Method (SISM). The minimum energy path was computed for a simultaneous reversal of two nanomagnets. The full trajectory of the energy landscape as a function of step number and temperature is shown in Fig. S2 [SM]. (b) Temporal evolution of the electron temperature $T_e$, phonon temperature $T_{ph}$, average magnetization modulus $\langle |M| \rangle$, and average component of the magnetization along the $x$-direction $\langle M_x \rangle$ following excitation of the artificial square ice shown in (c), initially in a Type II vertex configuration, with a 60 fs laser pulse at high fluence (7 mJ/cm$^2$), obtained from Landau-Lifshitz-Bloch (LLB) simulations. $\langle |M| \rangle$ is normalized to its equilibrium value at 300 K and $\langle M_x \rangle$ is normalized to its initial value at $t = 0$. The quantity $\langle|M|\rangle$ reflects the average local magnetization amplitude and tracks the evolution of the spin temperature, while $\langle M_x \rangle$ corresponds to the net magnetization along the x-direction and is sensitive to the global magnetic order. It can be seen that $\langle M_x \rangle$ does not fully recover after the pulse, indicating a transition from Type II to Type I vertex configurations. The corresponding results for the low-fluence case, in which $\langle M_x \rangle$ recovers and the vertices remain in the Type II configuration, are shown in Fig. S4 [SM]. (c) Snapshots from LLB simulations of the magnetization configuration at different time delays following ultrafast demagnetization, illustrating the emergence of multidomain vortex-like structures within the nanomagnets. The full sequence of magnetization snapshots from the LLB simulations, including additional intermediate time steps, is presented in Fig. S5 [SM].

# Supplemental Material

## I. Laser Exposure Procedure and Validation of Single-Pulse Reversal Mechanism

To ensure experimental reproducibility and minimize fluctuations in laser output, we exposed the artificial square ices to a continuous train of femtosecond pulses rather than a single-pulse excitation. This is because we found that manually triggering the laser to emit a single pulse led to significant pulse-to-pulse energy variability, which compromised the reliability of the measurements. Therefore, in all experiments discussed in the main text, the artificial square ices were exposed to the laser for 1 s at a repetition rate of 200 kHz. Nevertheless, control experiments confirmed that the formation of Type I vertex domains can indeed occur following a single laser pulse, provided that the pulse energy is sufficiently high.

Despite using a continuous train of femtosecond pulses, the magnetic configuration observed after laser exposure reflects the cumulative average effect of independent switching events induced by each individual pulse. This is because the magnetization fully recovers between consecutive pulses, since they are separated by several microseconds, which is far longer than the nanosecond timescale required for ultrafast demagnetization and recovery [Fig. 1(d)].

Importantly, the laser spot has a Gaussian profile and, during each individual laser pulse, all nanomagnets located in the central region where the local fluence exceeds the switching threshold can reverse their magnetization within the same excitation-recovery event. The resulting vertex configuration therefore emerges from a sequence of discrete, single-pulse reversal processes, each acting on a recovered magnetic state.

Moreover, to further verify that the observed magnetic relaxation is driven by individual pulses and not by cumulative heating effects due to the high repetition rate, we repeated the same experiments at lower laser repetition rates down to 1 kHz. This allowed sufficient time for the system to cool between pulses, ensuring that there is no significant rise in average temperature while maintaining stable laser operation. The vertex populations obtained at lower repetition rates fall within the statistical uncertainty of those reported in the main text for a 200 kHz repetition rate, confirming that the system does not accumulate heat during high-repetition-rate exposure, and that the magnetic relaxation originates from individual laser-induced nanosecond-long events.

## II. Simplified and Improved String Method: Energy Barrier Calculations

To quantify the energy required for magnetization reversal from a Type II to a Type I vertex configuration in an artificial square ice, we performed numerical calculations of the minimum energy path (MEP) connecting the initial and final magnetic states using the Simplified and Improved String Method [40–42]. The method identifies the minimum energy transition path between two relaxed states by iteratively minimizing the total magnetic energy along a discretized trajectory in configuration space.

The simulations were carried out using the finite-difference micromagnetic solver magnum.np [43], following the modeling approach found in Koraltan *et al*. [40] (Model 3 – a full micromagnetic model). In our case, the system geometry consists of two unit cells of artificial square ice with periodic boundary conditions, capturing the essential dipolar interactions while minimizing boundary effects.

The system initially contains identical Type II vertices. We then evaluate the transition towards a configuration containing only Type I vertices.

The material parameters used in the simulation are consistent with experimental values for 8 nm-thick permalloy ($Ni_{80}Fe_{20}$) thin films: saturation magnetization $M_S$ = 830 kA/m, exchange stiffness A = 17 pJ/m, and Curie temperature $T_C$ = 770 K. Temperature effects are incorporated by rescaling $M_S$ and A as a function of temperature [40]. The energy barrier $\Delta E(T)$ given in Fig. 4(a) is the difference between the energy at the saddle point along the MEP, $E$, and the energy of the initial Type II vertex configuration, $E_0$ [Fig. S2]. Note that, while this saddle point appears as a local maximum along the minimum energy path, it is in fact a saddle point in the full configuration space, representing the lowest-energy transition state between the two vertex configurations.

We model the simultaneous reversal of two nanomagnets at a vertex, as this is the minimal correlated switching event that can transform a Type II into a Type I vertex, while preserving the ice rule. This choice is motivated by the ultrafast experimental conditions, where laser excitation acts as a quasi-instantaneous global stimulus. Given that thermal activation is only possible during a narrow high-temperature window (≲100 ps), sequential switching of two coupled nanomagnets is unlikely to occur within this limited timeframe. The full energy profiles along the MEP for multiple temperatures, obtained with the SISM, are given in Fig. S2. The energy barrier [Fig. 4(a)] corresponds to the highest point along the minimum energy path connecting the initial and final magnetic states. These profiles confirm that, although the saddle point lowers with increasing temperature, the thermal energy $k_B T$ (66 meV) remains insufficient to overcome the barrier until temperatures approach the Curie point (770 K). This is evident from the fact that, at lower temperatures, the barrier height significantly exceeds $k_B T$, making thermal activation improbable. Snapshots of some of the magnetization configurations along the discretized transition path for $T$ = 300 K are shown in Fig. S3, where the color scale indicates the in-plane magnetization component. These snapshots provide a representative view of the magnetization evolution during reversal along the MEP, which proceeds via a curling of the magnetization within the nanomagnet.

### III. Modeling of Ultrafast Magnetization Dynamics with LLB Micromagnetic Simulations

To model the ultrafast magnetization dynamics in artificial square ice, we employed micromagnetic simulations based on the Landau-Lifshitz-Bloch (LLB) equation, using the same parameters adopted for the energy barrier calculations. The LLB equation is coupled to the electron temperature obtained from the solution of the two-temperature model [47,48]. The simulated system, shown in Fig. S4(b), consists of two unit cells of an artificial square ice with periodic boundary conditions and a discretization of 2 nm. The electronic specific heat is defined as $C_e = \gamma_e T_e$, with $\gamma_e$ = 553 J/(m$^3$K$^2$), and the phonon specific heat is $C_{ph}$ = 2.07×10$^6$ J/(m$^3$K). The electron-phonon coupling constant is set to $G_{e-ph}$ = 4.05×10$^{18}$ J/(m$^3$K). We assume an optical absorption coefficient of 0.25, a pulse duration of 60 fs, and a characteristic heat diffusion time to the substrate of 50 ps.

The system is initialized in a uniform Type II configuration and relaxed at 300 K for 10 ns by integrating the LLB equation to minimize the magnetic energy. A single laser pulse is then applied. However, the fluence values used in these simulations are not directly comparable to those used experimentally. This is because the relevant physical quantity is the absorbed energy density, which depends on optical absorption, geometry, and substrate. As a result, the simulated fluence values should be regarded as qualitative indicators of sub- and above-threshold excitation regimes (referred to in the main text as low and high fluence, respectively, corresponding to 6 mJ/cm$^2$ and 7 mJ/cm$^2$).

The temporal evolution of the electron and phonon temperatures, $T_{e,\,low}$ and $T_{ph,\,low}$, respectively, as well as the average magnetization modulus ⟨|$M$|⟩$_{low}$, and the average component of the magnetization along the *x*-direction ⟨$M_x$⟩$_{low}$, following excitation with a fluence of 6 mJ/cm$^2$, is presented in Fig. S4(a). In this case, the electron temperature remains below the Curie temperature, and the system only partially demagnetizes, retaining approximately 20% of its initial magnetization at 500 fs. Magnetization subsequently recovers to ≈ 70% of its initial value by 10 ps, and the initial Type II configuration is restored.

A different behavior is observed at higher fluence (7 mJ/cm$^2$), as shown in Fig. 4(b) and (c) of the main text and in the extended snapshots of Fig. S5. Here, the system is fully demagnetized and remains in a disordered state until ≈ 4 ps. At this point in time, the average magnetization modulus ⟨|$M$|⟩ begins to recover, and dipolar interactions become effective. Nucleation of magnetized regions occurs within the individual nanomagnets, resulting in a transient multidomain state at ≈ 100 ps. This state does not correspond to the ground-state configuration but rather reflects the presence of multiple nucleation sites embedded in a thermally disordered background. The competition between exchange and dipolar interactions leads to domain coarsening and finally a collapse into a single-domain configuration. During this relaxation, vortex-like textures may transiently form before the system ultimately stabilizes in a low-energy Type I configuration.

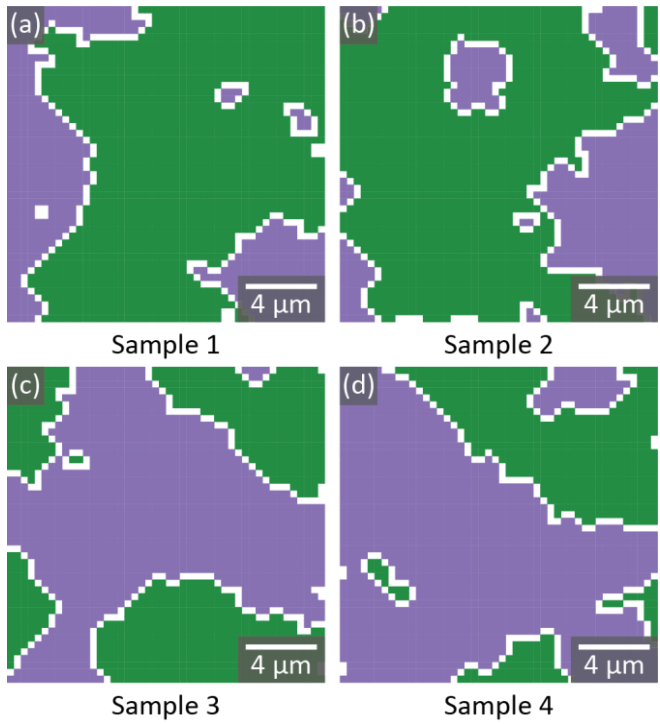

FIG. S1. (a)-(d) Simplified vertex configuration maps of the magnetic configuration of the artificial square ices (Samples 1-4, respectively) exposed to a decreasing fluence protocol over time, from 9.7 mJ/cm$^2$ to 6.1 mJ/cm$^2$. Here the Type I symbols are replaced with squares of the same color to highlight domains where the macrospins have the opposite sign. Type II and Type III symbols are omitted. The corresponding vertex populations are given in Fig. 3(b).

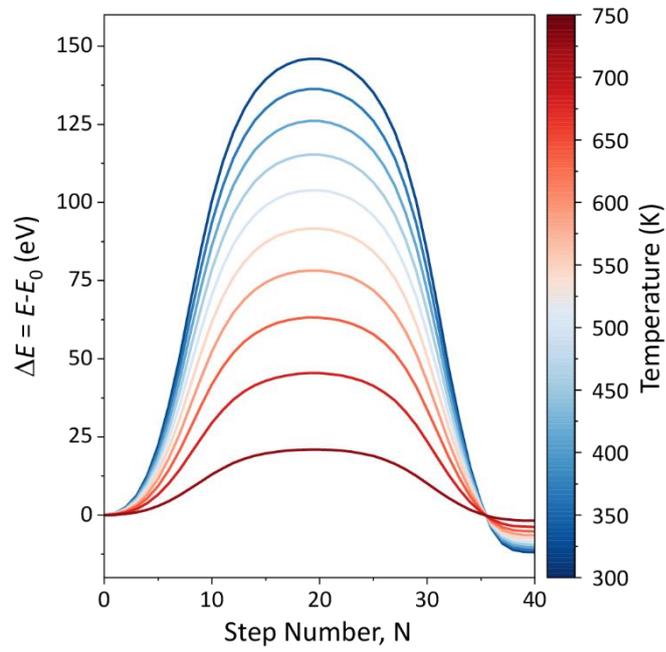

FIG. S2. Energy profiles along the minimum energy path (MEP) connecting a Type II and a Type I vertex configuration [Fig. S3], calculated using the Simplified and Improved String Method (SISM). The vertical axis gives the energy difference $\Delta E$ between the current energy of the state $E$ and the initial energy $E_0$ of the configuration where all vertices are in a Type II configuration, normalized for one vertex, while the horizontal axis denotes the discrete step index along the MEP.

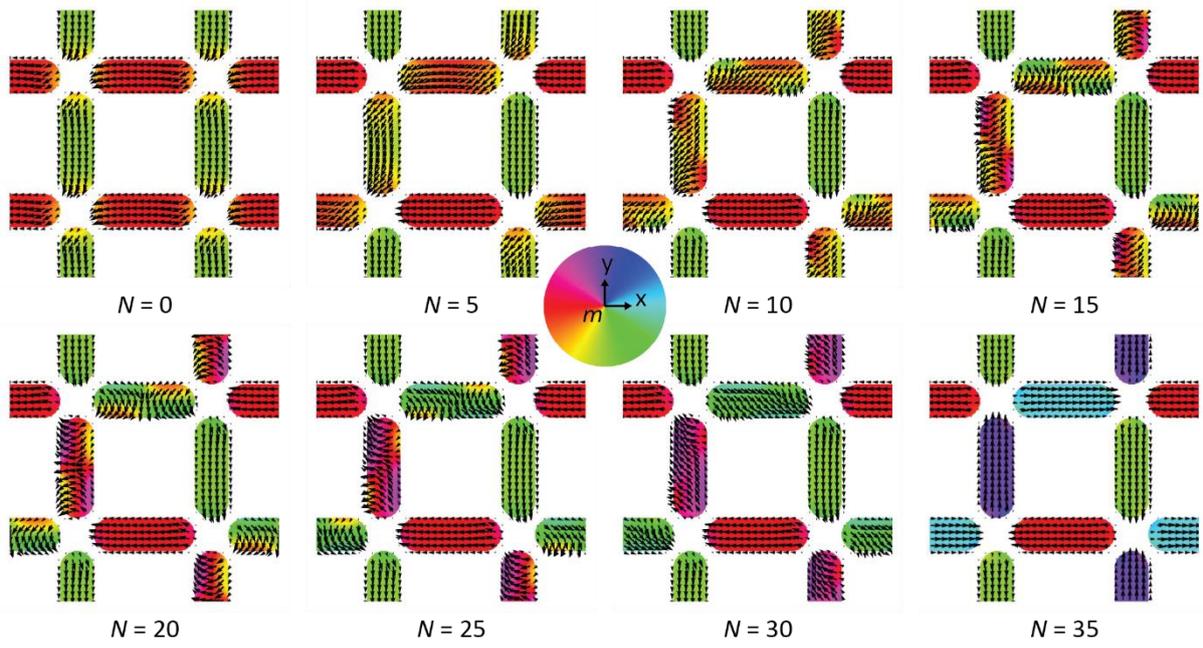

FIG. S3. Representative snapshots of the magnetization configurations at step $N$ of the discretized trajectory in configuration space along the minimum energy path connecting a Type II to a Type I vertex configuration for $T$ = 300 K, as obtained via the Simplified and Improved String Method (SISM). The arrows and the color scale represent the in-plane component of the magnetization.

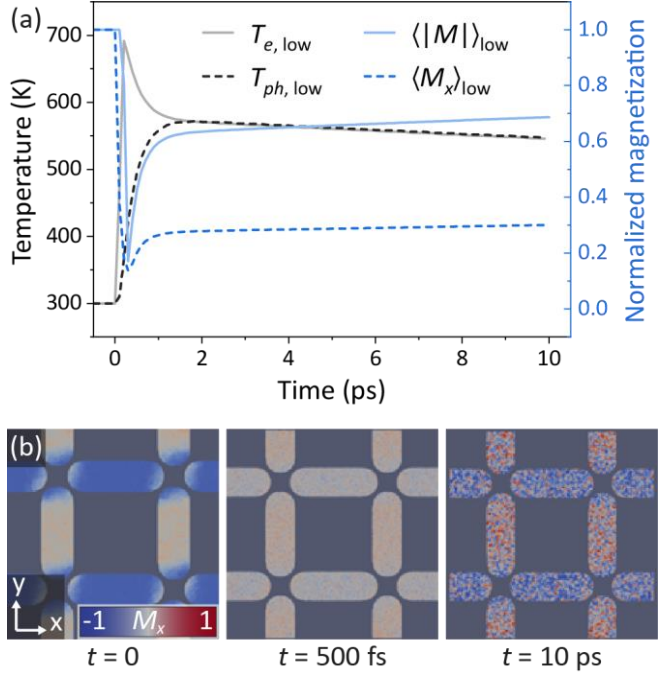

FIG. S4. Micromagnetic simulations of ultrafast magnetization dynamics. (a) Temporal evolution of the electron temperature $T_{e,\,low}$, phonon temperature $T_{ph,\,low}$, average magnetization modulus $\langle |M| \rangle_{low}$, and average component of the magnetization along the $x$-direction $\langle M_x \rangle_{low}$ following excitation of the artificial square ice shown in (b), initially in a Type II vertex configuration, with a 60 fs laser pulse at low fluence (6 mJ/cm$^2$), obtained from Landau-Lifshitz-Bloch (LLB) simulations. In this case, the electron temperature remains below the Curie temperature and $\langle M_x \rangle_{low}$ recovers after the pulse, indicating that the system returns to the initial Type II configuration. (b) Snapshots from LLB simulations of the magnetization configuration at different time delays following ultrafast demagnetization, illustrating the evolution of the magnetization texture within the nanomagnets. The fact that the system returns to its original magnetic configuration after excitation indicates that no irreversible changes occurred, and thus no transition to a lower-energy state took place.

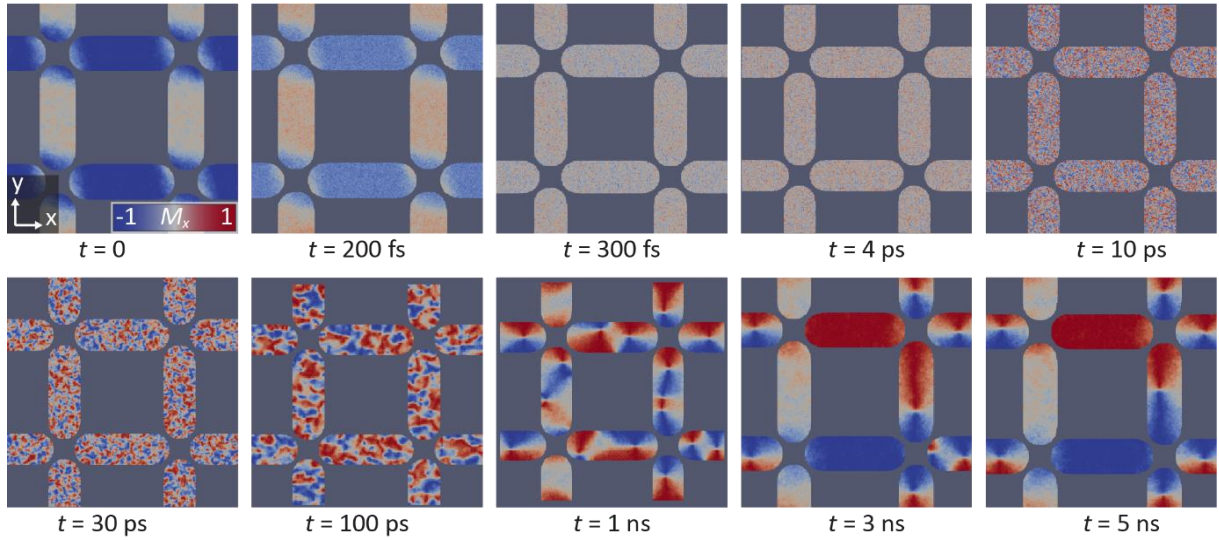

FIG. S5. Extended sequence of LLB simulation snapshots showing the evolution of the in-plane magnetization at successive time delays after high-fluence laser excitation (7 mJ/cm$^2$). The images illustrate the evolution from the initial Type II configuration to vortex-like intermediate states. While the full formation of low-energy Type I order is not observed within the simulated time window (up to 5 ns), the magnetization textures suggest a progressive reorganization towards such configurations.